\documentclass[graybox]{svmult}

\usepackage{mathptmx}       % selects Times Roman as basic font
\usepackage{helvet}         % selects Helvetica as sans-serif font
\usepackage{courier}        % selects Courier as typewriter font
\usepackage{type1cm}        % activate if the above 3 fonts are
\usepackage{makeidx}         % allows index generation
\usepackage{graphicx}        % standard LaTeX graphics tool
\usepackage{multicol}        % used for the two-column index
\usepackage[bottom]{footmisc}% places footnotes at page bottom

\makeindex             % used for the subject index
\begin{document}

\title*{ALMA polarimetric studies of rotating jet/disk systems}
% Use \titlerunning{Short Title} for an abbreviated version of
% your contribution title if the original one is too long
%\author{Name of First Author and Name of Second Author}
\author{F. Bacciotti, 
J.M. Girart, 
M. Padovani,
L. Podio,
R. Paladino, 
L. Testi,
E. Bianchi,
D. Galli,
C. Codella, 
D. Coffey,
C.  Favre and 
D. Fedele
}
% Use \authorrunning{Short Title} for an abbreviated version of
% your contribution title if the original one is too long
\institute{Bacciotti, F. \at Istituto Nazionale di Astrofisica - Osservatorio Astrofisico di Arcetri,
Largo Enrico Fermi, 5, I-50125 Firenze, Italy \email{fran@arcetri.astro.it}
\and Girart, J.M., \at Institut de Ci\`encies de l'Espai (ICE, CSIC), Can Magrans, s/n, E-08193 Cerdanyola del Vallès, Catalonia
\and Paladino R., \at Istituto Nazionale di Astrofisica - Istituto di Radioastronomia
Via P. Gobetti, 101 40129 Bologna, Italy
\and Testi, L., \at European Southern Observatory, Karl-Schwarzschild-Strasse 2, 85748 
Garching bei M\"unchen, Germany
\and Bianchi, E., \at Institut de Plan\'etologie et d'Astrophysique de Grenoble (IPAG)
Universit\'e Grenoble Alpes, CS 40700, 38058 Grenoble C\'edex 9, France
\and Padovani, M. Podio, L., Galli, D., Codella, C. 
Favre, F. Fedele, D. 
\at Istituto Nazionale di Astrofisica - Osservatorio Astrofisico di Arcetri,
Largo Enrico Fermi, 5, I-50125 Firenze, Italy
\and Coffey, D., \at School of Physics, University College Dublin,
Belfield, Dublin 4, Ireland
}
%
% Use the package "url.sty" to avoid
% problems with special characters
% used in your e-mail or web address
%
\maketitle

%\abstract*{Each chapter should be preceded by an abstract (10--15 lines long) that summarizes the content. The abstract will appear \textit{online} at \url{www.SpringerLink.com} and be available with unrestricted access. This allows unregistered users to read the abstract as a teaser for the complete chapter. As a general rule the abstracts will not appear in the printed version of your book unless it is the style of your particular book or that of the series to which your book belongs.
%Please use the 'starred' version of the new Springer \texttt{abstract} command for typesetting the text of the online abstracts (cf. source file of this chapter template \texttt{abstract}) and include them with the source files of your manuscript. Use the plain \texttt{abstract} command if the abstract is also to appear in the printed version of the book.}

\abstract{We have recently obtained polarimetric data at mm wavelengths with ALMA for the young systems DG Tau and CW Tau,
for which the rotation properties of jet and disk have been investigated in previous high angular resolution studies.   
The motivation was to test the models of magneto-centrifugal launch of jets via the determination of the magnetic configuration at the disk surface.   The analysis of these data, however, reveals that self-scattering of dust thermal radiation  dominates the polarization pattern.
It is shown that even if no information on the  magnetic field can be derived in this case, the polarization data are a powerful tool for the diagnostics of the properties and the evolution of dust in protoplanetary disks.}

%%%%%%%%%%%%%%%%%%%%%%%%%%%%%%%%%%%%%%%%%%%%%%%%%%%%%%%%%%%%%%%%%%%%%%%%%%
\section{Introduction}
\label{sec:1}
%%%%%%%%%%%%%%%%%%%%%%%%%%%%%%%%%%%%%%%%%%%%%%%%%%%%%%%%%%%%%%%%%%%%%%%%%%
The process of formation of stars and planets is one of the most intriguing topics in current astrophysics. In recent years, high angular resolution studies like the ones conducted with the Hubble Space Telescope (HST) and the   Atacama Large Millimeter/submillimeter Array (ALMA) have allowed us to advance significantly in the knowledge of protoplanetary disks and associated outflows. In particular, the sensitivity of ALMA allowed us to study the physical properties of young systems with an unprecedent combination of spectral and spatial resolution. One of the principal  aims of such studies is to correctly set the initial conditions for planet formation. 

In this context, the determination of the disk magnetic configuration   
is of particular interest, as magnetic fields may be responsible for the 
extraction of the excess of angular momentum from the system. 
This can occur via  magneto-rotational instabilities generating an effective 
viscosity for horizonthal transport \cite{Balbus1991}.
Such instabilities, however, have proven to be ineffective for disk realistic conditions 
\cite{BaiStone2013}. 
Alternatively, protostellar jets generated by  magneto-centrifugal acceleration
can transport angular momentum vertically along the ordered 
strong magnetic field attached to the star and the disk (\cite{Pudritz07, Frank14}).
Magneto-centrifugal winds can originate from the star ('stellar winds'), the disk  
co-rotation radius ('X-winds') or from a wider range of disk radii (extended
'disk winds'). 
In any case proto-planetary disks are expected to be strongly magnetized, 
which would have fundamental implications for planetary formation and migration models 
\cite{Turner14}. 

The increase in sensitivity in  mm-wave polarimetry 
has opened a new possibility to  investigate  the disk properties, and in particular the magnetic configuration. 
Polarimetry, in fact, has long been believed to provide the orientation  
of  magnetic field lines, as non-spherical dust grains tend to align with their short axis perpendicular to the direction of the magnetic field ('grain alignment'), giving rise to linear polarization of the emission
\cite{Andersson2015}. 
%Therefore, 
%one expectes a radial orientation of the polarization E vectors in a disk with   
%predominantly toroidal  magnetic field.

Polarization, however, can also arise  from self-scattering of the thermal emission 
of dust grains of size of the order of the radiation wavelength. 
In this case the  models show that  the polarization direction is parallel to the minor axis 
for inclined disks \cite{Yang2017, Kataoka2017}.

A third effect producing polarized emission is the alignement of non-spherical grains with 
an anisotropic radiation field. For a centrally illuminated disk, the linear polarization would present for this mechanism  a circular pattern centred on the source \cite{Tazaki2017}. 

%============================================================================================
\begin{figure}
%\sidecaption[t]
% Use the relevant command for your figure-insertion program
% to insert the figure file.
% For example, with the option graphics use
\includegraphics[scale=.4]{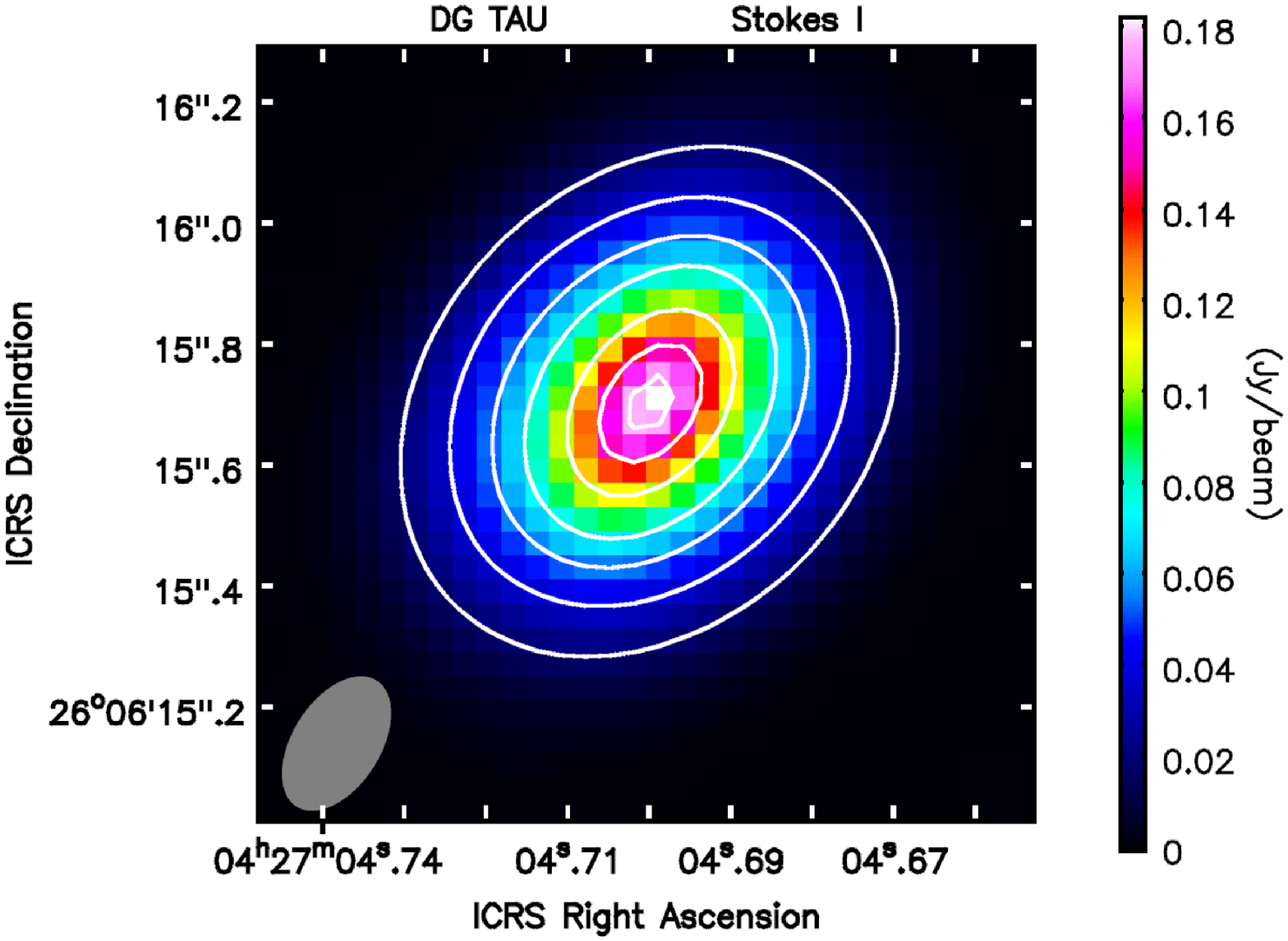}
\includegraphics[scale=.4]{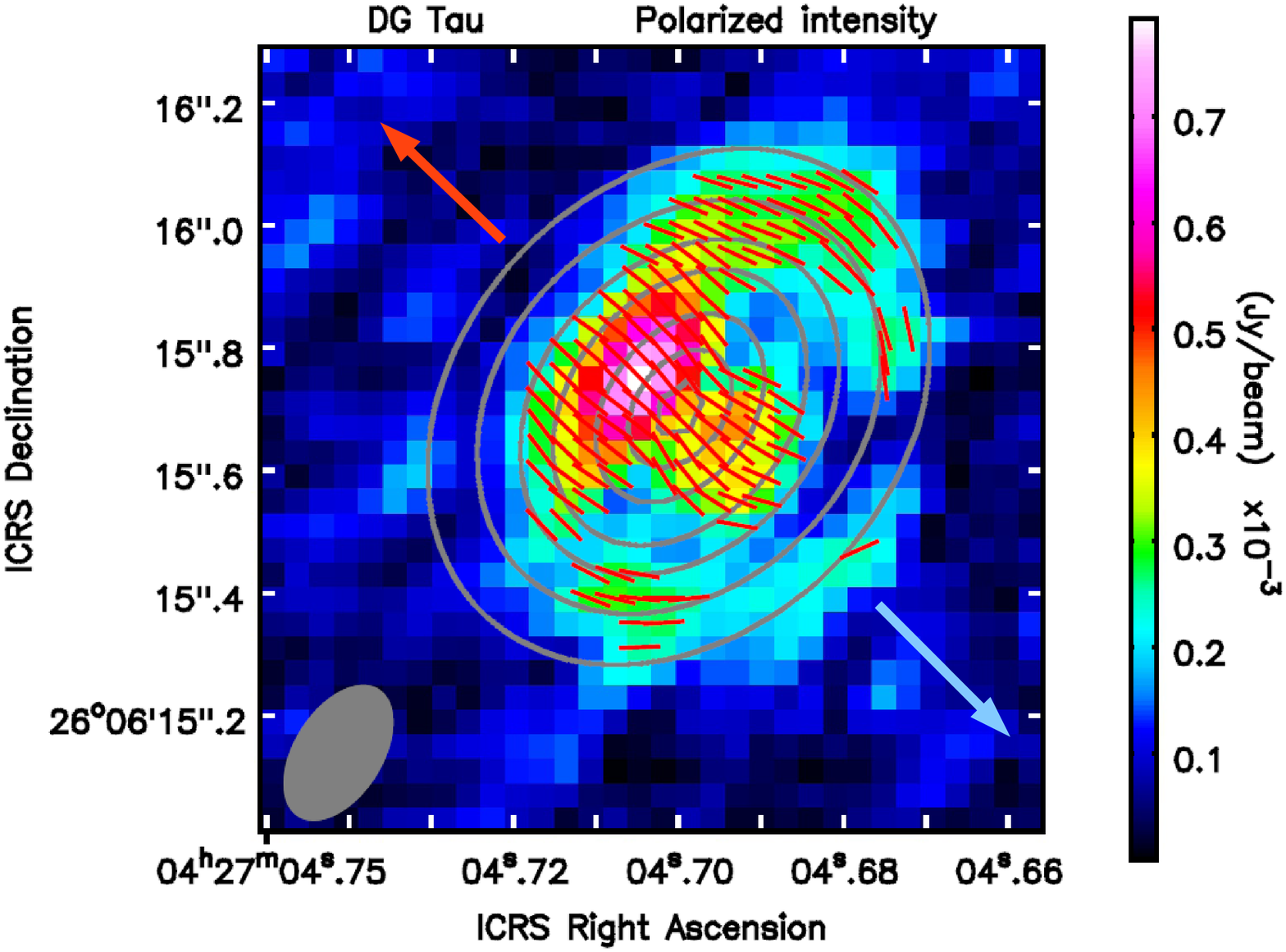}
%
% If no graphics program available, insert a blank space i.e. use
%\picplace{5cm}{2cm} % Give the correct figure height and width in cm
%
\caption{
{\em Top}
Total emission map at 870 $\mu$m 
in the disks around DG Tau.
Contour levels are [0.1,~0.2,~0.3,~0.4,~0.6,~0.8,~0.95] $\times$ peak value, 
which is 182.4 mJy beam$^{-1}$ for DG Tau.
{\em Bottom}
Linearly polarized intensity $P$. Contours are as in top panel and the
Polarization angle is indicated with vector bars.  
The arrows indicate the jet orientation, with the   
disk near-side on the same side of the red arrow.  As discussed in Sect.\ref{sec:3}, 
the evident asymmetry of the polarized intensity indicates a flared geometry for the disk. 
}
%\caption{If the width of the figure is less than 7.8 cm use the \texttt{sidecapion} command to flush the caption on the left side of the page. If the figure is positioned at the top of the page, align the sidecaption with the top of the figure -- to achieve this you simply need to use the optional argument \texttt{[t]} with the \texttt{sidecaption} command}
\label{fig:1}       % Give a unique label
\end{figure}
%===================================================================================================

The first studies on  protostellar envelopes
allowed  the identification  of large-scale  hourglass-shaped 
and twisted patterns consistent with the winding-up of magnetic field lines 
due to the rotation of the envelope
\cite{Girart2006, Rao2009}. 
Subsequent  studies at moderate resolution 
reported detections of polarized emission on protostellar discs 
(e.g. \cite{Rao2014, Kataoka2016a}) but in none of these cases the 
polarization structure showed a clear relationship with the expected
magnetic configuration.
More recently, the inner disk scales  have been 
reached thanks to the advent of ALMA. Maps of the polarization of the dust 
emission have been derived for various targets, with resolution
down to 0.''1 - 0.''2  (12 - 15 AU in nearby forming systems).
A polarisation level of 0.5 - 2\% can easily be detected over the
relevant regions of the disk  with the expected ALMA sensitivity.
These studies show that all the mechanisms mentioned above can produce 
polarization, but  dust self-scattering appears to be dominant \cite{Stephens2017, 
Kataoka2017, Lee2018, Girart2018, Hull2018, Alves2018}.

In this framework, we have started a project to map the polarization 
properties of young evolved Class II systems 
with associated  jets, as these systems  offer fundamental observational constraints in 
both the cases in which scattering or magnetic properties dominate the polarized emission.
In particular, we are interested in the sources for which the rotation kinematics 
of both the jet and the disk has been studied. The determination of the magnetic 
configuration for these targets  
can  constitute the ultimate proof of the validity of the magneto-centrifugal 
mechanism for the launch of jets, which can realize the exctraction of the excess angular momentum. 

Following this line of research  we selected and recently observed 
with ALMA the  Class II sources DG Tau and CW Tau. These targets 
have been the subject of numerous studies in the past, both for their disks  
and their collimated jets 
\cite{Isella2010, Pietu2014, Bacciotti2002, Hartigan2004}. 
The two sources  are nearby (d$\sim$140 pc), 
free of their parental envelope, are oriented favourably for polarization measurements
and their dust emission is sufficiently strong.  Importantly, the known kinematics of the jets 
and disks allows one to identify immediately the near-side of the disks and to give constraints on the expected magnetic configuration.

Despite our expectations, 
the observed polarization  (shown in Sect.~\ref{sec:2})  
does not seem to convey a simple interpretation in terms of an ordered  magnetic structure. Instead,
the results are fully consistent with self-scattering of the dust emission (see  discussion in Sect.~\ref{sec:3}). 
This allows us to derive from the polarization measurements 
new information on the size and distribution of dust grains in the disk \cite{Bacciotti18}. 

%Use the template \emph{chapter.tex} together with the Springer document class SVMono (monograph-type books) or SVMult (edited books) to style the various elements of your chapter content in the Springer layout.

%Instead of simply listing headings of different levels we recommend to
%let every heading be followed by at least a short passage of text.
%Further on please use the \LaTeX\ automatism for all your
%cross-references and citations. And please note that the first line of
%text that follows a heading is not indented, whereas the first lines of
%all subsequent paragraphs are.
%============================================================================================
\begin{figure}
%\sidecaption[t]
% Use the relevant command for your figure-insertion program
% to insert the figure file.
% For example, with the option graphics use
\includegraphics[scale=.45]{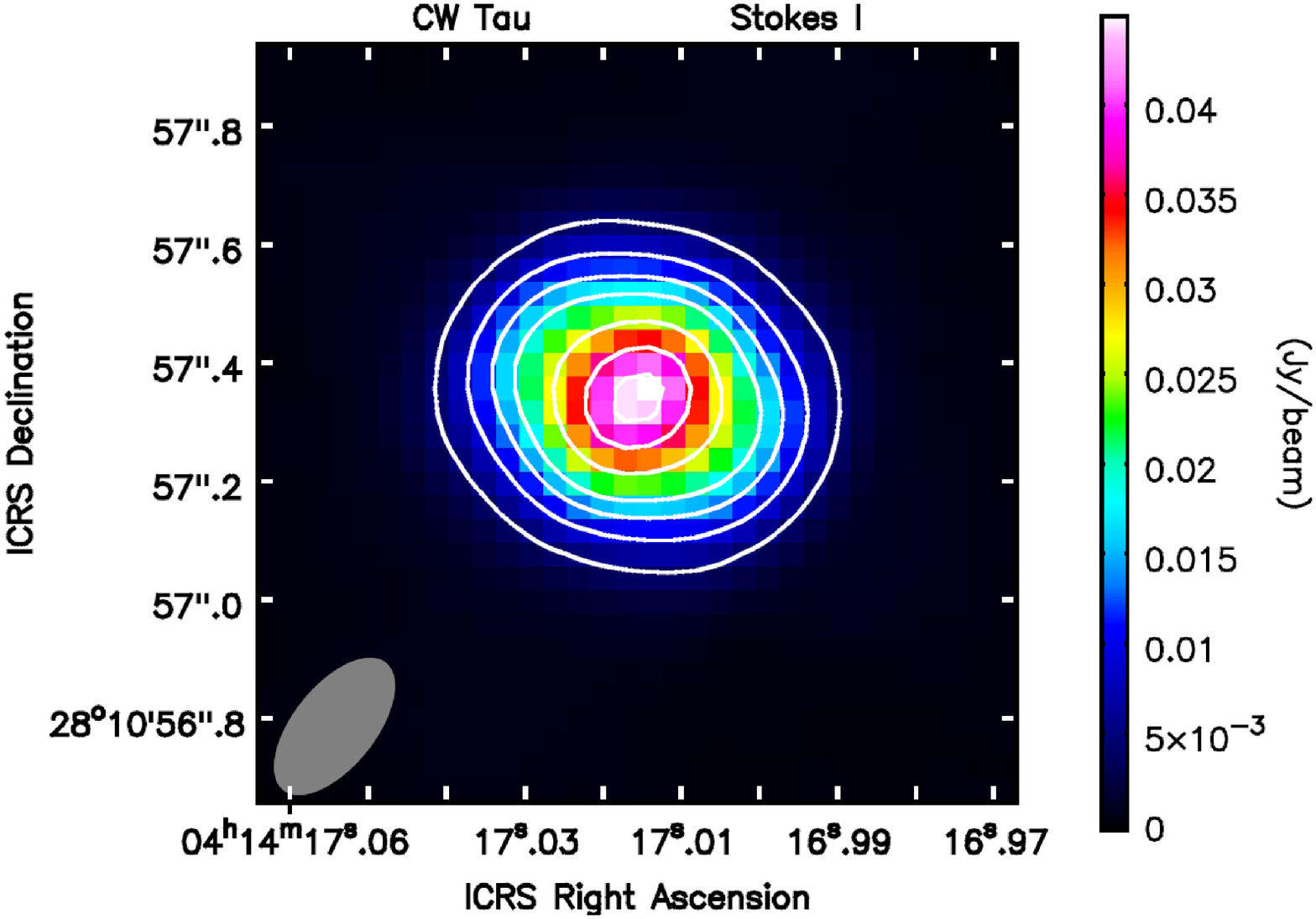}
\includegraphics[scale=.45]{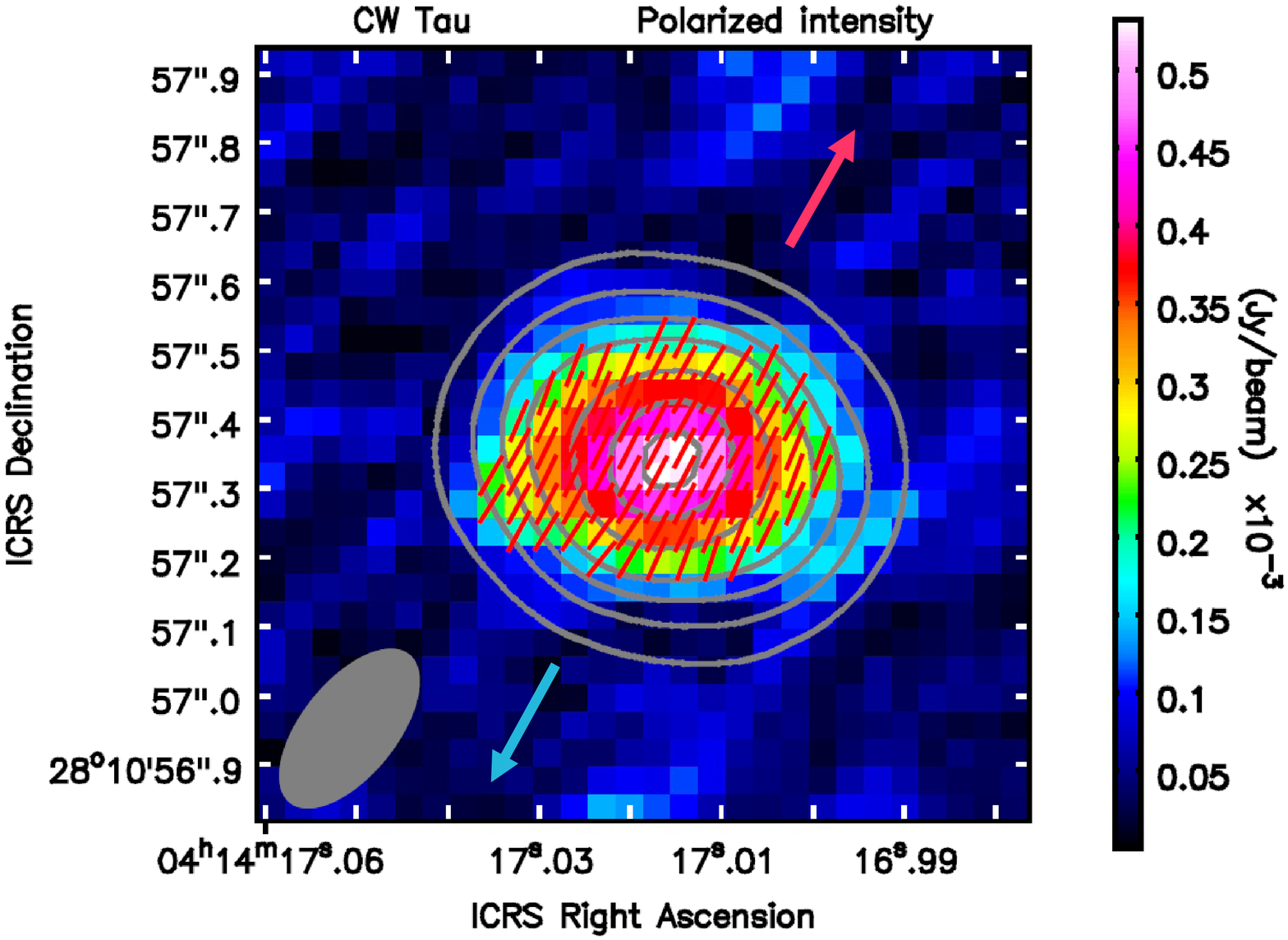}
%
% If no graphics program available, insert a blank space i.e. use
%\picplace{5cm}{2cm} % Give the correct figure height and width in cm
%
\caption{
{\em Top}
Total emission map at 870 $\mu$m 
in the disks around CW Tau.
Contour levels are [0.1,~0.2,~0.3,~0.4,~0.6,~0.8,~0.95] $\times$ peak value, 
which is 44.9 mJy beam$^{-1}$ for CW Tau.
{\em Bottom}
Linearly polarized intensity $P$. Contours are as in top panel. 
polarization angle, $\chi$, is indicated with fixed-length vector bars.  
The arrows follow the jet orientation.  
Disk near-side lies towards the receding jet lobe (red arrow).
The central symmetry of the polarized intensity suggests a geometrically thin disk (see Sect.\ref{sec:3}).
}
%\caption{If the width of the figure is less than 7.8 cm use the \texttt{sidecapion} command to flush the caption on the left side of the page. If the figure is positioned at the top of the page, align the sidecaption with the top of the figure -- to achieve this you simply need to use the optional argument \texttt{[t]} with the \texttt{sidecaption} command}
\label{fig:2}       % Give a unique label
\end{figure}
%=================================================================================================

\section{Polarized emission from DG Tau and CW Tau}
\label{sec:2}

In the following we illustrate 
the main featurs emerged from the polarimetric observations of DG Tau and CW Tau. 
More details can be found in \cite{Bacciotti18}.

The two targets were observed in full polarization mode  
%(simbad RA 4h 14m 17.0s, $\delta$ 28$^{\circ}$ 10' 57.4")
in July 2017 within the ALMA Cycle 3 in Band 7 (870 $\mu$m). 
%The spectral setup included four spectral windows, 1.875 GHz wide, centred at 
%the standard ALMA band 7 polarization frequencies (336, 338, 348 and 350 GHz).
%The spectral resolution was 55 km s$^{-1}$.
%Two successful executions 
%as part of the session scheme 
%were made on 2017 July 11, 
The configuration of the interferometer for these observations 
included 40 antennas,  giving an angular resolution of about 0.$''$2.
The datasets were reduced and analysed with the Common Astronomical Software
Application (CASA) software. 
From the Stokes I, U, Q maps we obtain the linear polarization intensity, 
$P=\sqrt{Q^2 + U^2 }$, the linear polarization fraction, $p=P/I$, 
and the polarization angle, $\chi=0.5 \arctan (U/Q)$, 
i.e. the direction of polarization of the electric field.  
%The ALMA instrumental error is reported to 
%be  0.1\% on $p$, and at least $2^{\circ}$ on $\chi$ 
%(ALMA polarization casaguide). 
%\vspace{0.5cm}

\subsection{ DG Tau }

The case of DG Tau is illustrated in Figure~\ref{fig:1}. 
The integrated flux is 880.2 $\pm$ 9.4 mJy,                      
% 0.880 con IMSTAT 864.2 com IMFIT 
with peak intensity of 182.4 $\pm$ 1.4 mJy beam$^{-1}$,                    
%182.37 with IMSTAT 152.6 con IMFIT -  IMSTAT RMS 15.12
The FWHM along the major and minor axis are  0.$''$45 and  0.$''$36, respectively. 
These values imply $i_{\rm disk} \sim 37^{\circ}$, while the measured disk PA is  135.4$^{\circ} \pm {2.5^\circ}$, almost perpendicular to PA$_{\rm jet}$ = 46$^{\circ}$. 

The polarization properties of DG Tau are illustrated in 
the bottom panel of Figure~\ref{fig:1}.
The near-side of the disk is brighter in polarized intensity, with the peak emission  on the minor axis, displaced by $\sim$ 0.$''$07 from the photocenter of the total intensity. 
This feature, observed here for the first time in a disk around 
a low mass star, indicates that the disk has a flared geometry \cite{Yang2017}.
In the outer disk region, between 0.$''$3
and 0.$''$5 from the source, the polarized emission is distributed
in a belt-like structure of lower intensity.  
The polarization vectors are nearly aligned with the disk minor axis in the central region, 
while they change orientation and become more azimuthal beyond 0.$''$3.
The linear polarization fraction (not shown) reflects the distribution of the polarized intensity. Averaging  
over the whole disk area one finds $p =$ 0.41 $\pm$ 0.17\%.

\subsection{CW Tau }
Figure~\ref{fig:2}, top panel, illustrates the total intensity of the 870 $\mu$m continuum emission in CW Tau. 
For CW Tau, the integrated flux is 145.1 $\pm$ 1.4 mJy,                    
with peak intensity of 44.9 $\pm$ 0.3 mJy beam$^{-1}$.
The FWHM along the major and minor axis is  0.$''$35 and 0.$''$18 respectively. 
From these values, we estimate a disk inclination $i_{\rm disk}$ with respect to the line of sight of $\sim 59^{\circ}$. The disk PA is 60.7$^{\circ} \pm {1.9^\circ}$, almost 
perpendicular to the jet PA$_{\rm jet}=$-$29^{\circ}$.

The bottom panel of Figure~\ref{fig:2} provides the  map of the linearly polarized intensity $P$.
This is centrally peaked and does not show any significant asymmetry. 
The polarization vectors 
%(parallel to the radiation electric field) 
are very well aligned along the minor axis of the disk. 
The polarization fraction, $p$, turns out to be almost constant 
in the disk central region of the disk, and it is on average 1.15 $\pm$ 0.26 \%.

\section{Dust properties derived from polarization measurements}
\label{sec:3}

The observed  polarization properties can be explained in both sources in terms of 
self-scattering of the thermal dust emission \cite{Kataoka2015,Yang2017}. This is in agreement with the findings in other protoplanetary disks  \cite{Kataoka2017, Girart2018, Hull2018}. 
If self-scattering dominates the polarization, no information can be gathered on the orientation of the magnetic field. This unfortunate occurence, however, is partly compensated by the fact that
the comparison of the observed maps and the predictions of the models for self-scattering give
constraints on the  size of th dust  population of particles  and on the geometry the disk.  

\subsection{Grain Size} 
\begin{figure}
%\sidecaption[t]
% Use the relevant command for your figure-insertion program
% to insert the figure file.
% For example, with the graphicx style use
\includegraphics[scale=.4]{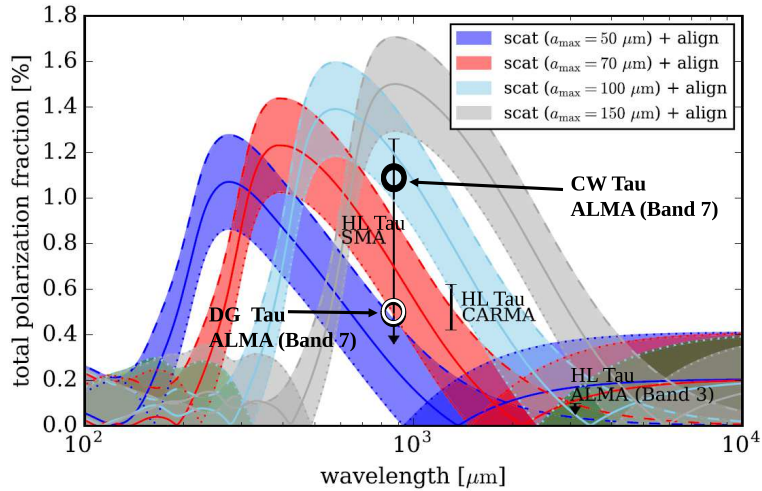}
%
% If no graphics program available, insert a blank space i.e. use
%\picplace{5cm}{2cm} % Give the correct figure height and width in cm
%
\caption{
Diagnostics of grain size using a diagram adapted from \cite{Kataoka2017}.
The vertical bars refer to past observations for the disk around HL Tau with variuos interferometers.
Our observations, performed at 870 $\mu$m in ALMA Band 7, are indicated by the  
the black and white circles. Their position correspond to 
the average polarization fraction we measured for CW Tau and DG Tau, respectively.
The average grain size turns out to be larger for CW Tau (about 100 $mu$m) than for DG Tau (50 - 70 $\mu$m).
}
%\caption{If the width of the figure is less than 7.8 cm use the \texttt{sidecapion} command to flush the caption on the left side of the page. If the figure is positioned at the top of the page, align the sidecaption with the top of the figure -- to achieve this you simply need to use the optional argument \texttt{[t]} with the \texttt{sidecaption} command}
\label{fig:3}       % Give a unique label
\end{figure}

The analysis  in \cite{Kataoka2015} indicates  that the maximum  grain size contributing to polarization from self-scattering 
at a given wavelength, $\lambda$, is comparable to $\lambda/2\pi$. This implies  a distribution peaked 
around 140 $\mu$m in our case. 
Further constraints can come from the diagnostic diagrams that correlate  grain size, 
wavelength and polarization fraction, as investigated, e.g. in \cite{Kataoka2017}.  
Using our values of the average polarization fraction and the diagrams in  
this work  we estimate that 
the maximum grain size giving rise to the observed polarization is  
in the range 50 - 70 $\mu$m for DG Tau and about 100  $\mu$m for CW Tau (see Fig.\,\ref{fig:3}).

%=====================================================================================
% For figures use
%
%=====================================================================================

\subsection{Grain settling toward the disk midplane}
In CW Tau, the distribution of the polarized intensity is symmetric, and polarization vectors are nearly parallel to the 
minor axis, with no curvature toward the outer  disk.
Since the disk around CW tau is reported to be optically thick
by \cite{Pietu2014}, the polarization models indicate that  
the observed  features are consistent with self-scattering from a geometrically thin disk. 
Thus our observations indicate that the relatively larger grains in CW tau (see Section above)  
are more settled close to the disk midplane. 

On the contrary, in DG Tau the polarization angle alignment is accompanied
by an asymmetry in polarization intensity. This combination is consistent with the expectations of models of self-scattering in disks 
of intermediate or high optical depth, and with a finite angular thickness \cite{Yang2017}.  
%%%%%%%%%%%%%%%%%%%%%%%%%
The disk is  moderately optically thick according to \cite{Isella2010}, and the polarization maps are
in agreement 
with the model expectations of \cite{Yang2017}. 
Thus the observed asymmetry indicates that the scattering 
grains have not yet settled to the midplane. 

\subsection{Hints for substructures in DG Tau ?}
 We now consider the outer region of the DG Tau disk, i.e. beyond 0.$''$3  from the source. The bottom panel of Figure\, \ref{fig:1} shows structures in the polarized emission which do not correspond to any feature in the total intensity at the same resolution. In addition, a change in the orientation of the polarization pattern is observed. 
A possible explanation may come from a drop in the optical depth at 0.$''$3 from the star, corresponding to about 45 au at the distance of the system. This may imply that there is a substructure in the disk density  at this location, like a gap or a ring, not revealed in the total emission (see discussion in \cite{Bacciotti18}). The nature of such a structure has still to be revealed, and will require higher angular resolution observations. We anticipate, however, that a recent study in the emission of molecular lines  has revealed a ring in the emission of formaldheide (H2CO)
whose inner border is at 0.''3 from the star, coincident with the change in the polarization properties \cite{Podio18}. Thus it appears that polarization maps nicely complement the investigations in total emission. 
  
\section{Conclusions}
The ALMA observations of disks are providing new and precious information for the understanding of the formation of planets around young stars. 
In particular, the window opened recently by polarimetric capabilities allows us to set important constraints on the distribution and  early evolution of the dust component of disks. 
As other systems recently observed, DG Tau and CW Tau show polarization properties at 870 $\mu$m
dominated by  self-scattering of the dust thermal emission. Overall, DG Tau appears to be in 
a less evolved state than CW Tau. In addition, structural peculiarities  are revealed by the polarized emission. Our  analysis thus indicates that polarimetry will be a powerful tool in the studies of the evolution of protoplanetary disks.

\begin{acknowledgement}
FB wishes to dedicate this work to the loving memory of her mother-in-law 
Giovanna (Janet) Pesenti. FB also  wish thanks the editors for their understanding 
in a difficult time. Support is acknowledged from the project 
EU-FP7-JETSET (MRTN-CT-2004-005592).
\end{acknowledgement}

\end{document}